\documentclass{ws-ijmpa}

\begin{document}

\markboth{Weeks} {Detecting Topology}

\catchline{}{}{}

\title{Detecting Topology in a Nearly Flat Hyperbolic Universe}

\author{\footnotesize Jeffrey R. Weeks}

\address{15 Farmer Street\\
Canton NY 13617-1120, U.S.A.}

\maketitle

\pub{Received (Day Month Year)}{Revised (Day Month Year)}

\begin{abstract}
Cosmic microwave background data shows the observable universe to
be nearly flat, but leaves open the question of whether it is
simply or multiply connected.  Several authors have investigated
whether the topology of a multiply connect hyperbolic universe
would be detectable when $0.9 < \Omega < 1$.  However, the
possibility of detecting a given topology varies depending on the
location of the observer within the space.  Recent studies have
assumed the observer sits at a favorable location.  The present
paper extends that work to consider observers at all points in the
space, and (for given values of $\Omega_m$ and $\Omega_\Lambda$
and a given topology) computes the probability that a randomly
placed observer could detect the topology. The computations show
that when $\Omega = 0.98$ a randomly placed observer has a
reasonable chance ($\sim 50\%$) of detecting a hyperbolic
topology, but when $\Omega = 0.99$ the chances are low ($< 10\%$)
and decrease still further as $\Omega$ approaches one.

\end{abstract}

\section{Introduction}

Analysis of recent cosmic microwave background (CMB) data suggests
an approximately flat universe with the total energy density
parameter $\Omega$ almost surely lying in the range $0.9 < \Omega
< 1.1$.\cite{sievers,netterfield}  The near flatness of the
observable universe does not preclude a multiconnected spatial
topology, but may push the topology to a scale larger than the
horizon radius, making it difficult or impossible to detect.
Recent studies have examined the extent to which a nontrivial
topology may or may not be observable in a locally spherical
universe\cite{wlu02} or a locally hyperbolic
one.\cite{gomero1,aurichsteiner,inoue3,gomero2} In a locally flat
universe there is no {\it a priori} relationship between the
topology scale and the horizon scale, so a great deal of luck
would be required for the two to coincide.

In their most recent study of multiconnected hyperbolic
universes,\cite{gomero2} Gomero, Rebou{\c c}as, and Tavakol
examine the seven smallest known hyperbolic topologies and find
that for a set of cosmological parameters given by Bond {\it et
al.}\cite{bond} with $\Omega = 0.99$, five of the seven topologies
would be potentially detectable using CMB methods, while for a set
of parameters given by Jaffe {\it et al.}\cite{jaffe} with $\Omega
= 0.98$, all seven would be potentially detectable.

A topology is considered potentially detectable by an observer at
a point $p$ in the space if the observer's horizon
radius\footnote{One may substitute the last scattering surface at
$z \simeq 1100$ for the absolute horizon at $z = \infty$ with
little effect on the horizon radius.} $r_{hor}$ exceeds the
injectivity radius\footnote{Given a point $p$ in a multiconnected
space, the {\it injectivity radius} at $p$ is defined to be the
radius of the largest ball centered at $p$ whose interior does not
overlap itself.  Equivalently, twice the injectivity radius is the
length of the shortest topologically nontrivial path from $p$ to
itself.} $r_{inj}(p)$ at $p$.  In a 3-torus and in many spherical
topologies, the injectivity radius $r_{inj}$ is constant
throughout the whole space, so if the topology is potentially
detectable by an observer at point $p$, it is detectable by any
other observer at any other point $q$ in the same space. In
hyperbolic topologies, by contrast, the injectivity radius varies
from point to point, so a hyperbolic topology might be detectable
by an observer at point $p$ (where the injectivity radius
$r_{inj}(p)$ is small), but undetectable by a different observer
at some other point $q$ (where the injectivity radius $r_{inj}(q)$
is large).

In their most recent work,\cite{gomero2} Gomero, Rebou{\c c}as,
and Tavakol consider the question of detectability from the most
favorable point in the space, that is, from a point of minimal
injectivity radius.  The present article extends their work by
considering the detectability of the topology at arbitrary
locations in the space. The variation of the injectivity radius
across the space will be summarized in an {\it injectivity
profile} showing what fraction of the space's volume has a given
injectivity radius.  Combining the injectivity profile
(Section~\ref{SectionInjectivityProfiles}) with an estimate for
the horizon radius (Section~\ref{SectionHorizonRadius}) reveals
the probability that a randomly placed observer could potentially
detect the topology, that is, it tells the fraction of the
manifold's volume in which $r_{hor} > r_{inj}$
(Section~\ref{SectionResults}).

\section{Injectivity Profiles}
\label{SectionInjectivityProfiles}

\subsection{Definition}

An {\it injectivity profile} is a histogram showing how much of a
manifold's volume has a given injectivity radius.  For example, in
the simple histogram in Fig.~1 the first bar shows that the
injectivity radius lies in the range [0.2, 0.3] for 10\% of the
manifold's volume;  the second bar shows that the injectivity
radius lies in the range [0.3, 0.4] for 20\% of the manifold's
volume;  and so on. In the studies presented below, each histogram
will have 240 bins of width 0.0025, spanning the range of
injectivity radii from 0 to 0.6.  In the limit, as the bin width
goes to zero, the assignment of a finite volume percentage $\Delta
V/V$ to each bin of finite width $\Delta r$ is replaced by a
limiting density distribution $(d V/V) / d r$ as shown in Fig.~2.
The density distribution's discontinuities reflect a preferred set
of short closed geodesics and may be the subject of a future
paper, but nevertheless hold no importance for the present study.

\begin{figure}
\centerline{\psfig{file=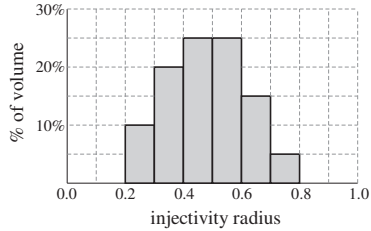,width=5cm}}
\vspace*{8pt} \caption{An injectivity profile shows how much
  of a manifold's volume has a given injectivity radius.}
\label{FigureSampleHistogram} 
\end{figure}

\begin{figure}
\centerline{\psfig{file=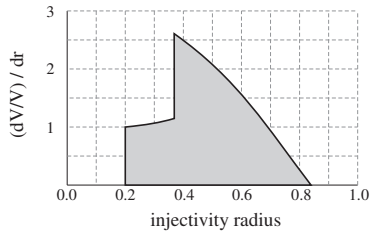,width=5cm}}
\vspace*{8pt} \caption{As the bin width goes to zero, the
  histogram becomes a density distribution.}
\label{FigureSampleDensity} 
\end{figure}

\subsection{Computational Overview}
\label{SubsectionInjectivityProfileComputation}

To construct an injectivity profile, begin with a fundamental
domain for the manifold.  The present study used a Dirichlet
domain computed by the computer program SnapPea.\cite{SnapPea}
SnapPea presents the Dirichlet domain in the Klein model of
hyperbolic space, where it looks just like an ordinary Euclidean
polyhedron. Cover the Klein model with a grid of points.  The grid
is rectangular relative to the Euclidean geometry of the model,
although of course it's nonrectangular relative to the intrinsic
hyperbolic geometry.  The present study used a $200 \times 200
\times 200$ grid.  The program scanned the grid, first rejecting
grid points lying on or beyond the sphere-at-infinity in the Klein
model, and then rejecting points lying outside the Dirichlet
domain.  For each grid point lying within the Dirichlet domain,
the program computed the injectivity radius at that point
(Subsection~\ref{SubsectionInjectivityRadius}) to determine the
correct bin in the histogram, then computed the true hyperbolic
volume of the surrounding grid cell
(Subsection~\ref{SubsectionGridCellVolume}) and added that volume
to the correct bin.  After all grid points were processed, the
program printed the volume in each bin to a file.

\subsection{Computing the Injectivity Radius}
\label{SubsectionInjectivityRadius}

At first glance, computing the injectivity radius at a point $p$
is trivially easy:  just apply elements $\gamma$ of the holonomy
group $\Gamma$ to $p$ and see what the minimum translation
distance $dist(p, \gamma(p))$ is.\footnote{The holonomy group
$\Gamma$ is also known as the group of covering transformations.
We follow the convention of saying {\it holonomy group} when
thinking of $\Gamma$ geometrically as a group of isometries, but
saying {\it group of covering transformations} when thinking of
$\Gamma$ topologically as a group of homeomorphisms.} The
injectivity radius will be half that minimum:  $r_{inj}(p) =
\min_{\gamma\in\Gamma}\frac{dist(p, \gamma(p))}{2}$.  The only
problem is that $\Gamma$ is an infinite group, so we must decide
ahead of time how many elements---and which elements---of $\Gamma$
to apply.

Our selection criterion for choosing a finite subset of $\Gamma$
depends on the concept of an isometry $\gamma$'s {\it basepoint
translation distance}, defined to be the distance $dist(O,
\gamma(O))$ that $\gamma$ translates the origin $O$. Say we want
to find all $\gamma \in \Gamma$ that translate a given point $p$ a
distance at most $\ell$.  Any such isometry $\gamma$ has basepoint
translation distance $dist(O, \gamma(O)) \leq dist(O, p) + dist(p,
\gamma(p)) + dist(\gamma(p), \gamma(O)) \leq D_{out} + \ell +
D_{out}$, where $D_{out}$ is the outradius of the Dirichlet domain
$D$.  In other words, if we consider all isometries $\gamma$ with
basepoint translation distance at most $\ell + 2 D_{out}$, we are
sure to find all translates of $p$ closer than the distance
$\ell$.

How large must $\ell$ be?  A simple argument\footnote{Let $\gamma$
be the face pairing isometry taking one of the Dirichlet domain's
faces closest to the origin $O$ to its mate.  Because the chosen
face is maximally close to the origin, $dist(O, \gamma(O)) = 2
D_{in}$, where $D_{in}$ is the inradius of the Dirichlet domain
$D$. Thus for any point $p$ in the Dirichlet domain, $dist(p,
\gamma(p)) \leq dist(p, O) + dist(O, \gamma(O)) + dist(\gamma(O),
\gamma(p)) \leq D_{out} + 2 D_{in} + D_{out}$.} shows that every
point $p$ has a translate $\gamma(p)$ lying at a distance less
than $2 D_{in} + 2 D_{out}$.  So if we were to choose $\ell$ in
the preceding paragraph to be $2 D_{in} + 2 D_{out}$, we would be
guaranteed to find the nearest translate of every point $p$, and
would therefore know the injectivity radius $r_{inj}(p)$. This
method, while rigorous, is inefficient: the number of isometries
$\gamma$ with basepoint translation distance at most $\ell$ grows
exponentially in $\ell$, and the majority of those isometries are
unneeded.  A more efficient algorithm is simply to guess a
plausible value of $\ell$, find all isometries $\gamma$ with
basepoint translation distance at most $\ell + 2 D_{out}$, and
verify afterwards that for each point $p$ we found a translate
$\gamma(p)$ at a distance $dist(p, \gamma(p))$ less than $\ell$.
There is no guarantee that such a translate will always be found,
but if one is found then we rigorously know the minimum value of
$dist(p, \gamma(p))$ and hence the injectivity radius $r_{inj}(p)
=  dist(p, \gamma(p)) / 2 $.  In the present study a value of
$\ell = 1.3$ worked in all cases.

Given a Dirichlet domain $D$ and its face pairings, how do we find
all isometries $\gamma \in \Gamma$ with basepoint translation
distance $dist(O, \gamma(O))$ at most $\ell$?  A simple recursion
does the job:  start with the the original Dirichlet domain $D$;
add its immediate neighbors in the universal covering space, which
we think of as the translates $\gamma(D)$ for each of $D$'s face
pairing isometries $\gamma$;  then add the neighbors' neighbors,
which correspond to products $\gamma_2\gamma_1(D)$ of pairs of
face pairing isometries;  and so on.  Continue recursively,
keeping those images $\gamma(D)$ with $dist(O, \gamma(O)) \leq
\ell$ and discarding those with $dist(O, \gamma(O)) > \ell$.
Proposition 3.1 of \cite{HodgsonWeeks} shows that this algorithm
finds all isometries with $dist(O, \gamma(O)) \leq \ell$;  that
is, we needn't worry that the recursion will terminate when it
encounters only unwanted translates with $dist(O, \gamma(O)) >
\ell$, while some wanted translate with $dist(O, \gamma(O)) \leq
\ell$ remains hidden beyond them.

\subsection{Computing the Volume of a Grid Cell}
\label{SubsectionGridCellVolume}

Inscribe the Klein model of hyperbolic space in a cube with
corners at $(\pm 1, \pm 1, \pm 1)$, and cover the cube with a $200
\times 200 \times 200$ grid.  Each grid cell is then a small cube
of side length $10^{-2}$ and Euclidean volume $10^{-6}$.  The
hyperbolic volume represented by each grid cell differs from the
apparent Euclidean volume, and varies from cell to cell.

It's easy to compute the ratio of the true hyperbolic volume to
the apparent Euclidean volume.  The grid cells are small enough
that the volume ratio doesn't vary much within a given cell, so
pick any point $(x,y,z)$ within the cell, and let $r = \sqrt{x^2 +
y^2 + z^2}$ be its distance from the origin.  By symmetry the
volume ratio depends only on $r$, not on the particular point
$(x,y,z)$. So take a spherical shell of radius $r$ and thickness
$d r$, and compare its true hyperbolic volume to its apparent
Euclidean volume.  The apparent Euclidean volume is
\begin{equation}
V_E = 4 \pi r^2 d r.
\label{EqnEuclideanVolume}
\end{equation}
To compute the shell's true hyperbolic volume, project from the
Klein model onto the hyperboloid model of hyperbolic space
(Fig.~3).  Similar triangles show that the hyperbolic radius
$\rho$ has $\cosh\rho = 1 / \sqrt{1 - r^2}$ and $\sinh\rho = r /
\sqrt{1 - r^2}$. Thus the area of the sphere is $4 \pi \sinh^2\rho
= 4\pi r^2 / (1 - r^2)$.  To find its thickness, note that
$\tanh\rho = r$ so $d\rho = \cosh^2\rho\,d r = d r / (1 - r^2)$.
Hence the hyperbolic volume of the spherical shell is
\begin{equation}
\label{EqnVolumeRatio}
V_H = \frac{4\pi r^2 d r}{(1 - r^2)^2}
\label{EqnHyperbolicVolume}
\end{equation}
and the required ratio is
\begin{equation}
\frac{V_H}{V_E} = \frac{4\pi r^2 d r / (1 - r^2)^2}{4 \pi r^2 d r}
= \frac{1}{(1 - r^2)^2}.
\end{equation}

\begin{figure}
\centerline{\psfig{file=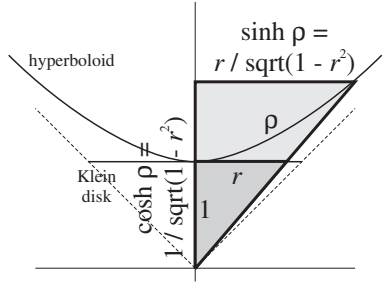,width=5cm}} \vspace*{8pt}
\caption{The Klein disk projects radially onto the hyperboloid
model of hyperbolic space, taking a sphere of Euclidean radius $r$
in the Klein disk to a sphere of intrinsic hyperbolic radius
$\rho$.}
\label{FigureVolumeRatio} 
\end{figure}

\section{Horizon Radius}
\label{SectionHorizonRadius}

The variable $r_{hor}$ represents the horizon radius {\it in
comoving coordinates}, that is, in units of the curvature radius
of the ambient hyperbolic space.  In a standard
Friedmann-Lema\^itre model with matter density today $\Omega_m$
and cosmological constant $\Omega_\Lambda$, one may compute
$r_{hor}$ by integrating the path of a photon backwards from $z =
0$ to $z = \infty$, obtaining
\begin{equation}
 r_{hor}=
   \sqrt{|\Omega_m+\Omega_\Lambda-1|}
   \int_0^\infty\frac{d z}
   {\sqrt{\Omega_\Lambda+(1-\Omega_m-\Omega_\Lambda)(z+1)^2
   +\Omega_{m}(z+1)^3}}.
 \label{EqnHorizonRadius}
\end{equation}
Replacing the upper limit of $z=\infty$ with the redshift $z=1100$
of last scatter would have little effect on the result.

For comparison with Gomero {\it et al.},\cite{gomero2} the
remainder of the present article will consider two sets of
plausible choices for the density parameters:
\begin{itemlist}
 \item $\Omega_m = 0.37$ and $\Omega_\Lambda = 0.61$, for which
       $\Omega_{total} = 0.98$ and $r_{hor} = 0.43$, and
 \item $\Omega_m = 0.37$ and $\Omega_\Lambda = 0.62$, for which
       $\Omega_{total} = 0.99$ and $r_{hor} = 0.30$.
\end{itemlist}

\section{Results}
\label{SectionResults}

A computer program, written in the C programming language and
freely available from {\tt
ftp://ftp.northnet.org/weeks/NearlyFlatHyperbolic}, computes
injectivity profiles using the algorithm outlined in
Subsection~\ref{SubsectionInjectivityProfileComputation} along
with the details explained in
Subsections~\ref{SubsectionInjectivityRadius} and
\ref{SubsectionGridCellVolume}.  For the first ten low-volume
hyperbolic manifolds from the Hodgson-Weeks
census,\cite{HodgsonWeeks} which include the seven manifolds
considered by Gomero {\it et al.},\cite{gomero2} the program
obtains the injectivity profiles shown in Fig.~4.  Shaded
backgrounds mark the cutoffs $r_{hor} = 0.43$ and $r_{hor} =
0.30$, corresponding to $\Omega_{total} = 0.98$ and
$\Omega_{total} = 0.99$, respectively. Only in the fraction of the
manifold to left of each cutoff would the topology be detectable.
Table~1 summarizes the results.

\begin{table}[h]
 \tbl{The fraction of each manifold's volume in which its topology
 is potentially detectable when $\Omega_{total} = 0.98$ or $0.99$.}
 {\begin{tabular}{@{}cccc@{}}
 \toprule  & manifold & $\Omega_{total} = 0.98$ & $\Omega_{total} = 0.99$ \\
 \colrule
  1 & m003(-3, 1) & 78\% &  4\% \\
  2 & m003(-2, 3) & 55\% &  2\% \\
  3 & m007( 3, 1) & 12\% &  0\% \\
  4 & m003(-4, 3) & 69\% &  4\% \\
  5 & m004( 6, 1) & 70\% & 16\% \\
  6 & m004( 1, 2) & 32\% &  5\% \\
  7 & m009( 4, 1) &  7\% &  0\% \\
  8 & m003(-3, 4) & 25\% &  9\% \\
  9 & m003(-4, 1) & 41\% &  8\% \\
 10 & m004( 3, 2) & 39\% &  5\% \\
 \botrule
\end{tabular}}
\end{table}

\begin{figure}
\centerline{\psfig{file=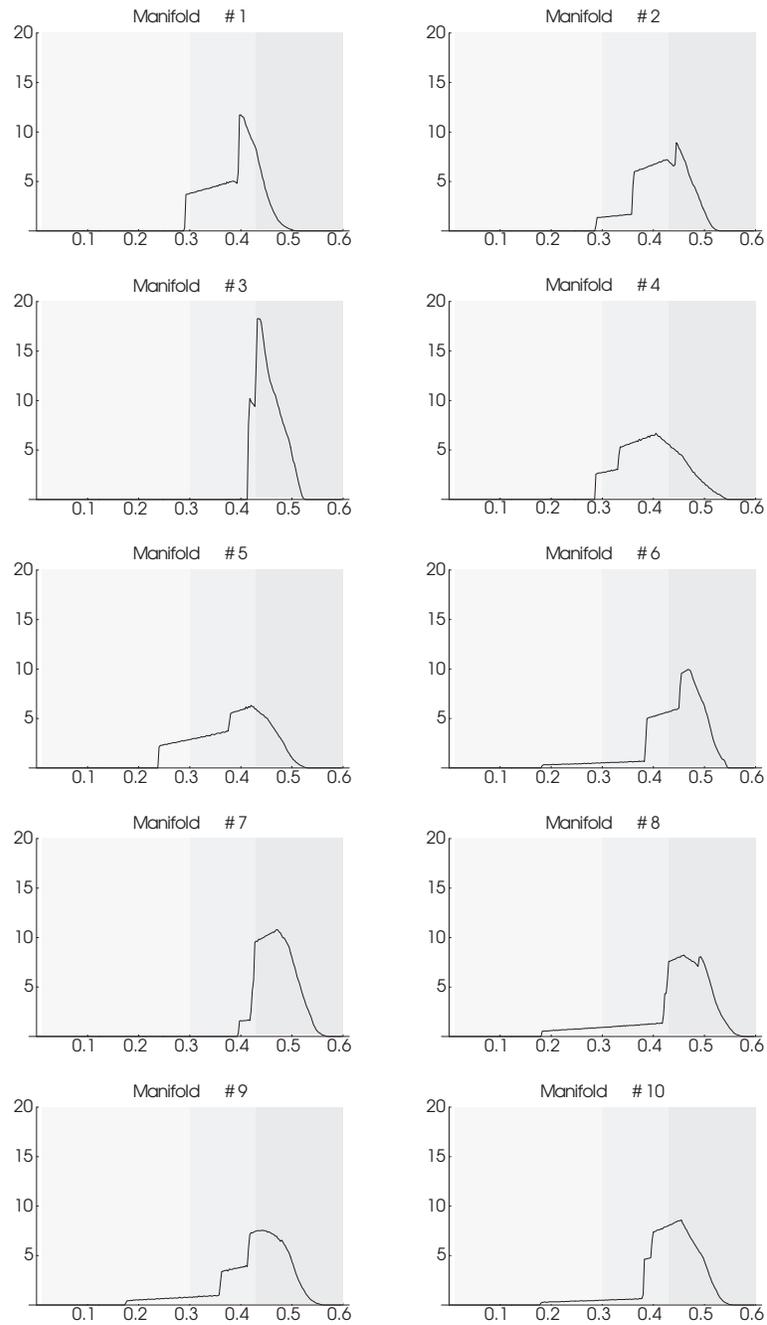,width=10cm}}
\vspace*{8pt} \caption{Injectivity profiles for the manifolds of
Table~1.}
\label{FigureResultsClosed} 
\end{figure}

\vspace{2cm}

 The manifolds in Table~1 contain no closed geodesics
shorter than 0.36.  However, there exist families of hyperbolic
3-manifolds containing arbitrarily short closed geodesics.  The
manifolds within each family approach a limiting {\it cusped
manifold} which is, in effect, a manifold with a geodesic of
length zero. Furthermore, the injectivity profiles of the
manifolds in each family approach the injectivity profile of the
limiting cusped manifold.  Fig.~5 shows the injectivity profiles
for the two smallest limiting cusped manifolds, {\it m003} and
{\it m004},\footnote{The cusped manifolds {\it m000}, {\it m001},
and {\it m002} are nonorientable and not the limit of any sequence
of closed manifolds.} while Table~2 shows the fraction of each in
which an observer could potentially detect the topology.  Note
that for $\Omega_{total}$ close to one, the fraction of the
manifold in which the topology is detectable is roughly
proportional to $1 - \Omega_{total}$.  The Dirichlet domains for
these cusped manifolds have infinite outradius, so in the
algorithm of Subsection~\ref{SubsectionInjectivityRadius} we
considered all isometries of basepoint translation distance less
than the plausible but arbitrary value of 2.5;  thus these results
for cusped manifolds are not rigorous, but they are most likely
correct nonetheless.  The beginning of each profile is noisy
because at most a few grid points fall into each bin;  refining
the grid diminishes this effect at the expense of a slower
computation.

\begin{table}[h]
 \tbl{The fraction of each cusped manifold's volume in which its topology
 is potentially detectable when $\Omega_{total}$ is close to one.}
 {\begin{tabular}{@{}ccccc@{}}
 \toprule manifold
 & $\Omega_{total} = 0.98$
 & $\Omega_{total} = 0.99$
 & $\Omega_{total} = 0.995$
 & $\Omega_{total} = 0.9975$ \\
 \colrule
 {\it m003} & 17\% &  8\% &  4\% &  2\% \\
 {\it m004} & 67\% & 32\% & 16\% &  8\% \\
 \botrule
\end{tabular}}
\end{table}

\begin{figure}
\centerline{\psfig{file=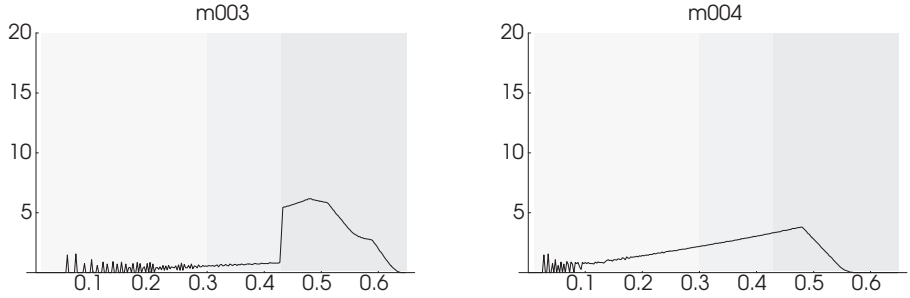,width=12cm}}
\vspace*{8pt} \caption{Injectivity profiles for the two smallest
limiting cusped manifolds.}
\label{FigureResultsCusped} 
\end{figure}

\vspace{2cm}

\section{Conclusions}

In the case $\Omega_{total} = 0.98$, the horizon radius $r_{hor} =
0.43$ is large enough that an observer at a random location in a
small hyperbolic universe would have a reasonable chance of
detecting the topology.  However, in the case $\Omega_{total} =
0.99$, the horizon radius $r_{hor}$ drops to $0.30$ and a random
observer would have little or no chance of detecting the topology.

There exist closed hyperbolic 3-manifolds with arbitrarily short
closed geodesics, so no matter how close $\Omega_{total}$ is to
one, there will always be infinitely many hyperbolic manifolds in
which well-placed observers could detect the topology. However,
the probability that an observer could detect the topology from a
random point in such a manifold decreases in proportion to $1 -
\Omega_{total}$.

\section*{Acknowledgements}

I thank the MacArthur Foundation for supporting my work.


\begin{thebibliography}{0}

\bibitem{sievers}
 J.L. Sievers {\em et al.}
 [{\tt arXiv:astro-ph/0205387}].

\bibitem{netterfield}
 C. Netterfield {\it et al.}
 [{\tt arXiv:astro-ph/0104460}].

\bibitem{wlu02}
 J. Weeks, R. Lehoucq and J.-P. Uzan
 submitted to {\it Class. Quant. Grav.}
 [{\tt arXiv:astro-ph/0209389}].

\bibitem{gomero1}
 G.I. Gomero, M.J. Rebou{\c c}as, and R. Tavakol,
 {\it Class. Quant. Grav.} {\bf18}, L145 (2001)
 [{\tt arXiv:gr-qc/0106044}].

\bibitem{aurichsteiner}
 R. Aurich and F. Steiner,
 {\it Mon. Not. R. Astron. Soc.} {\bf323}, 1016 (2001)
 [{\tt arXiv:astro-ph/0007264}].

\bibitem{inoue3}
 K.T. Inoue,
 {\it Prog. Theor. Phys.} {\bf106}, 39 (2001)
 [{\tt arXiv:astro-ph/0102222}].

\bibitem{gomero2}
 G.I. Gomero, M.J. Rebou{\c c}as, and R. Tavakol,
 {\it Int. J. Mod. Phys. A} {\bf A17} 4261 (2002)
 [{\tt arXiv:gr-qc/0210016}].

\bibitem{bond}
 J.R. Bond {\it et al.},
 in {\it Proc. of CAPP-2000 (AIP)}, CITA-2000-64.

\bibitem{jaffe}
 A.H. Jaffe {\it et al.}, {\it Phys. Rev. Lett.} {\bf86}, 3475 (2001).

\bibitem{SnapPea}
 J. Weeks, SnapPea:  a computer program for creating and studying
 hyperbolic 3-manifolds, available for free at {\tt www.northnet.org/weeks/SnapPea}.

\bibitem{HodgsonWeeks}
 C. Hodgson and J. Weeks,
 {\it Experimental Mathematics} {\bf3}, 261 (1994).

\end{thebibliography}
\end{document}